\documentclass[twocolumn,amsmath,amssymb,aps]{aastex62}
\usepackage[fleqn]{amsmath}
\usepackage{mathrsfs}

\usepackage{ctable}
\usepackage{graphicx}
\usepackage{dcolumn}
\usepackage{bm}
\usepackage{amsmath}

\received{}
\revised{}
\accepted{\today}
\submitjournal{The Planetary Science Journal}

\shorttitle{Formation conditions of Titan and Enceladus' building blocks}
\shortauthors{Anderson, Mousis, Ronnet}

\begin{document}

\title{Formation conditions of Titan and Enceladus' building blocks in Saturn's circumplanetary disk}

\email{sarah.anderson@univ-fcomte.fr}

\author[0000-0002-0786-7307]{Sarah E. Anderson}
\affil{Aix Marseille Univ, CNRS, CNES, LAM, Marseille, France}
\affiliation {Univ. Bourgogne Franche-Comt\'e, OSU THETA, Besan\c con, France}

\author{Olivier Mousis}
\affil{Aix Marseille Univ, CNRS, CNES, LAM, Marseille, France}

\author{Thomas Ronnet}
\affiliation{Lund Observatory, Department of Astronomy and Theoretical Physics, Lund University, Box 43, 22100, Lund, Sweden}

\begin{abstract}
The building blocks of Titan and Enceladus are believed to have formed in a late-stage circumplanetary disk around Saturn. Evaluating the evolution of the abundances of volatile species in this disk as a function of the migration, growth, and evaporation of icy grains is then of primary importance to assess the origin of the material that eventually formed these two moons. Here we use a simple prescription of Saturn's circumplanetary disk in which the location of the centrifugal radius is varied, to investigate the time evolution of the icelines of water ice, ammonia hydrate, methane clathrate, carbon monoxide and dinitrogen pure condensates. To match their compositional data, the building blocks of both moons would have had to form in a region of the circumplanetary disk situated between the icelines of carbon monoxide and dinitrogen at their outer limit, and the iceline of methane clathrate as their inner limit. We find that a source of dust at the location of centrifugal radius does not guarantee the replenishment of the disk in the volatiles assumed to be primordial in Titan and Enceladus. Only simulations assuming a centrifugal radius in the 66--100 Saturnian radii range allow for the formation and growth of solids with compositions consistent with those measured in Enceladus and Titan. The species are then able to evolve in solid forms in the system for longer periods of time, even reaching an equilibrium, thus favoring the formation of Titan and Enceladus building blocks in this region of the disk.
\end{abstract}

\keywords{planets and satellites: composition --- planets and satellites: formation --- planets and satellites: individual (Titan, Enceladus)}

\section{Introduction} \label{sec:intro}

The exploration of Saturn's satellite system by the {\it Cassini-Huygens} spacecraft has revealed several puzzling features regarding the compositions of the moons Titan and Enceladus, prompting the revision of their formation models. While the \textit{Huygens} probe's descent to Titan’s surface confirmed that the atmosphere is dominated by N$_2$ and CH$_4$, with a very low CO:CH$_4$ ratio ($\sim$10$^{-3}$) as previously observed by \textit{Voyager} and ground-based observations \citep{Ga97}, it also revealed a significant depletion of the primordial noble gases. The only definitively observed primordial noble gas detected by the Gas Chromatograph Mass Spectrometer (GCMS) aboard the Huygens probe was $^{36}$Ar, with a $^{36}$Ar/$^{14}$N ratio lower than the solar value by more than five orders of magnitude \citep{Ni05,Ni10}. The other primordial noble gases Kr and Xe (and $^{38}$Ar) were not detected by the GCMS instrument down to upper limits of 10$^{-8}$ relative to nitrogen \citep{Ni05}. These absences of detections are puzzling as noble gases are notable in the atmospheres of telluric planets \citep{Pe92,Wi02}, as well as in the atmosphere of Jupiter \citep{Ow99,Mo19}. The depletion in CO is also a strong constraint on Titan's composition since it is believed to have been more abundant than CH$_4$ in the protosolar nebula (PSN) \citep{Mu11,Bo04,Bi2017}. Since CO shares a similar volatility with N$_2$, its very low abundance in Titan's atmosphere is consistent with the strongly supported interpretation that the observed N$_2$ is probably not primordial and would result from photolysis, shock chemistry, or thermal decomposition of primordial NH$_3$ \citep{At78,Mc88,Ma07,Se11,Ma14,Mi19}. In addition, the flyby of Enceladus's south pole by the \textit{Cassini} spacecraft allowed the measurement of the composition of its plumes by the Ion and Neutral Mass Spectrometer (INMS), unveiling a 96--99\% concentration of H$_2$O, along with small amounts of CO$_2$ (0.3 to 0.8\%), CH$_4$ (0.1 to 0.3\%), NH$_3$ (0.4 to 1.3\%), and H$_2$ (0.4 to 1.4\%) \citep{Wa17}. Also, $^{36}$Ar, CO and N$_2$ were not detected by the INMS instrument in open source mode, strengthening the argument that the building blocks of Enceladus may have formed devoid of these three molecules.

Together, the measurements of Titan and Enceladus' compositions suggest they could have been assembled from similar building blocks. This scenario has been developed by \cite{Mo09a} and \cite{Mo09b} who proposed that both moons formed from building blocks initially produced in the protosolar nebula prior to having been partially devolatilized in a temperature range comprised between the formation temperatures of CO and N$_2$ pure condensates and the crystallization temperature of CH$_4$ clathrate in Saturn's circumplanetary disk (CPD). By doing so, Titan and Enceladus' building blocks would have been devoid of CO and N$_2$, while still keeping the entrapped CH$_4$ to match the observed compositions. However, these two studies did not investigate the transport (gas diffusion and drift of solid particles) of key volatiles around the locations of their respective icelines.

An iceline is defined as the radius at which the disk temperature is equal to the sublimation or condensation temperature of water-ice (or any species of interest) in the PSN and circumplanetary disks (hereafter CPDs). Inside the iceline, ice sublimates. Outside, ice remains stable, though the motion of particles within the disk would allow for solids to exist in front of this line as well as some vapors to exist beyond. Thus, the icelines induce the creation of peaks of abundances of volatile species in disks \citep{Al14,Mo19,Ag20}. Tracking the abundances of volatiles in both solid and vapor states around their respective icelines should then provide tighter constraints than previous works on the locations at which the building blocks of Titan and Enceladus formed, assuming that their volatile content is essentially primordial.

Here we aim to explore the time evolution of the icelines of H$_2$O ice, NH$_3$ hydrate, CH$_4$ clathrate, CO, and N$_2$ pure condensates,  i.e. the key volatiles needed to explain Titan and Enceladus' observed compositions, within Saturn's CPD and derive their impact on the formation conditions  of the two satellites' building blocks. Our approach is based on a simple prescription of a CPD model \citep{Ca2002,Sa10} and follows the dynamic radial evolution of icy dust and gases as they cross over the various icelines, estimating growth, fragmentation, and condensation as they drift within the disk. The simulation also follows the evolution of the different vapors as they condense on the grain surfaces or become enriched if icy grains evaporate.

An important parameter in our simulations is the location of the centrifugal radius $R_c$, which corresponds to the point where the angular momentum of the incoming gas is in balance with the gravitational potential of Saturn. It is also the injection point of the solid material entering the CPD from the PSN \citep{Ca2002,Sa10}, and its value is overall poorly constrained in CPDs. For instance, to account for the orbital distribution of the Galilean satellites, $R_c$ has been estimated to be in the $\sim$20--30~$R_{Jup}$ range in Jupiter's CPD \citep{Ru82}. This range could be somewhat larger if some inward type I migration of satellite embryos is taken into account, perhaps $R_c$ $\sim$35--40~$~R_{Jup}$ \citep{Ca2002}. In the case of Saturn's CPD, \cite{Sa10} have opted to set $R_c$~=~30~$R_{Sat}$, by similarity with previous studies of the Jovian CPD, while \cite{Ma09} finds $R_c$~=~66~$R_{Sat}$, by taking into account the specificity of the disk (lower mass of the planet and larger Hill's radius). On the other hand, recent simulations of gas accretion onto a CPD show that the material infalling towards the subdisk could be distributed out to much larger distances, of the order of $\sim$100 $R_p$ \citep{Su17}. Here, we investigate the influence of the variation of $R_c$'s location in the formation region of the satellites' building blocks. This allows us to show that the abundance of solid material within this disk strongly depends on the radial distance of its injection point.

Section \ref{sec:model} is dedicated to the description of the disk and transport models. Section \ref{sec:res} presents the results of several computations with different initial conditions for the delivery of particles from the CPD, namely different positions of $R_c$. Conclusions are presented in Sec \ref{sec:dis}.

\section{Model} \label{sec:model}

\subsection{Gas Starved Accretion Disk}
\label{sec:2.1} 

We consider a circum-Saturnian accretion disk with an inflow of material from the surrounding PSN based on the approaches of \cite{Ca2002} and \cite{Sa10}. It is limited by the outer radius $R_d$~=~200~$R_{Sat}$ by processes such as solar torques or through collisions with shocked regions. The CPD is fed through its upper layers from its inner edge up to the centrifugal radius $R_c$ by gas and gas-coupled solids inflowing from the PSN. In the following, we start with the assumption that $R_c$~=~30~$R_{Sat}$ \citep{Ca2002}, and also investigate several cases at different radii within the CPD (see Sec. \ref{sec:res}). In this 1D model, for each point of the radius $r$, we have an integrated surface density $\Sigma$ which describes the integrated density over the vertical slice of the disk. The quantities that interest us are the surface density of the dust $\Sigma_{d}$ and the surface density of the vapor $\Sigma_{v}$. These will be compared to the surface density of the PSN gas, $\Sigma_g$. Here, ``dust'' refers to icy grains. 

\begin{figure}[t]
\includegraphics[width=\linewidth]{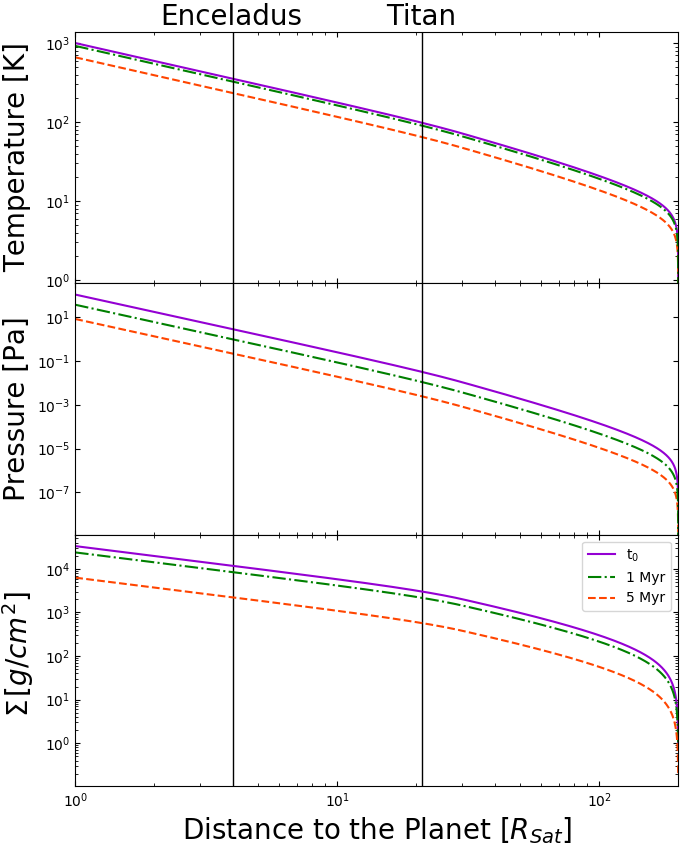}\label{fig1}
\caption{Disk steady-state pressure, temperature, and gas surface density for the gas starved accretion disk model of Saturn at $t$ = 0 (solid lines), $t$ = $10^6$ yr (dot-dashed lines), and $t$ = $5\times10^6$ yr (dashed lines) for $R_c$ = 30 $R_Sat$. The evolution of these quantities is very slow in the CPD and can even be seen as stationary over short timescales. The present day locations of Enceladus and Titan are shown for reference.}
\end{figure}



The surface density of the disk gas is given by \citep{Ca2002}:
\begin{equation}
\Sigma_g \simeq \frac{4 F_p}{15\pi\nu} \lambda(r),
\end{equation}

\noindent where $F_p$ is the total infall rate, $\nu$ is disk gas turbulent viscosity. The turbulent viscosity of the CPD is defined via $\nu$~=~$\alpha$$C_s^2$/$\Omega_{\rm K}$ \citep{Sh73}, where $\alpha$ (set to 10$^{-3}$) is the viscosity parameter. $\Omega_K$~=~$\sqrt{G M_{Sat} / r^3}$ is the Keplerian frequency with $G$ defined as the gravitational constant and $M_{Sat}$ is the mass of Saturn. $C_s$ is the isothermal sound speed given by $C_s~=~\sqrt{kT/\mu mp}$, where $k$ the Boltzmann constant, $\mu$ the mean molecular weight, and $m_p$ the proton mass. The coefficient $\lambda(r)$ is set equal to:

\begin{equation}
\begin{split}
\lambda(r) = \frac{5}{4}-\sqrt{\frac{R_c}{R_d}}-\frac{1}{4}\Big(\frac{r}{R_c}\Big)^2  ~{\rm for}~r<R_c, \\
{\rm and~} \lambda(r) = \sqrt{\frac{R_c}{r}}- \sqrt{\frac{R_c}{R_d}}    ~{\rm for}~r \ge R_c.
\end{split}
\end{equation}

In the steady accretion state, the total infall rate is regulated by a parameter $\tau_G~=~M_{Sat}/(dM/dt)^{-1}$, where $dM/dt$ is the inflow rate, so that $F_{p,0} = M_{Sat}/\tau_G$.  We adopt $\tau_G$ = 5 $\times$ 10$^6$ yr for Saturn \citep{Sa10}. The total infall rate follows an exponential law decay via $F_p~=~F_{p,0}\exp(-t/\tau_{\text{Dep}})$, with the timescale of CPD depletion $\tau_{Dep}$ set to 3$\times$ 10$^6$ yr \citep{Ca2006,Sa10}. 
      
The CPD is heated by luminosity from the central planet, viscous dissipation, and energy dissipation associated with the difference between the free-fall energy of the incoming gas and that of a Keplerian orbit. Assuming that viscous dissipation is dominant, the photosurface temperature of the CPD ($T_d$) is determined by a balance between viscous heating and blackbody radiation from the photosurface \citep{Ca2002,Sa10}:

\begin{equation}
\label{bal}
\sigma_{sb}T_d^4 \simeq \frac{9}{8}\Omega_K^2\nu\Sigma_g = \frac{3 \Omega_K^2}{10 \pi} F_p~\lambda(r)
\end{equation}

\noindent 
The disk temperature profile at $t$~=~0 can then be given as follows:

\begin{equation}
\label{temp}
T_{d,0}^4 = \frac{3 \Omega_K^2}{10 \pi \sigma_{sb}} F_{p,0}~\lambda(r) 
\end{equation}

\noindent Using this expression, the disk temperature can be written as a function of time:
\begin{equation}
T_{d} \simeq T_{d,0}\exp{\Big(\frac{-t}{4\tau_{dep}}\Big)}.
\end{equation}
        
The inflow rate is also constrained by the requirement that temperatures be low enough for pure condensates to remain stable in the outer regions of Saturn's CPD, a requirement which this model fulfills without alteration, showing similar values as \cite{Ca2002} for the Jupiter disk. Figure 1 presents the thermodynamic properties of the CPD model used throughout this paper with $R_c$ = 30 R$_{Sat}$.

\subsection{\label{sec:level2}Grain size evolution}

The dust evolution model is derived from the approach of \cite{Bi2012}. The particles are evenly injected into the disk with a uniform size $a~=~10^{-6}$ m. These micron-sized crystalline icy grains evolve in size and position through collisions, fragmentation, and radial drift due to gas drag. In our calculations, the size of grains increases before it reaches an equilibrium corresponding to the minimum value between fragmentation and radial drift.

Fragmentation occurs when the relative velocity of the dust grains due to turbulent motion exceeds the fragmentation velocity threshold $u_f$. We set the latter to 10 m s$^{-1}$ \citep{Bi2012}. The dust internal density is set to $\rho_s = 1 \text{g cm}^{-3}$. The size of the grains, limited by their fragmentation, is \citep{Bi2012}:

\begin{equation}
a_{\text{frag}} = 0.37\frac{2\Sigma_g}{3\pi\rho_s\alpha}\frac{u_f^2}{c_s^2}.
\end{equation}

\noindent In many cases, the grains drift inward too quickly for them to grow any larger. The size of these grains, limited by their drift, is given by:

\begin{equation}
a_{\text{drift}} = 0.55\frac{2 \Sigma_d}{\pi\rho_s}\frac{v_K^2}{c_s^2}\Big|\frac{\text{d}\ln P}{\text{d}\ln r}\Big|^{-1},
\end{equation}

\noindent where $v_K$ is the keplerian velocity, $P = c^2_s\rho_g$ and $\rho_g = \Sigma_g/2\pi H_g$ the pressure and gas density at midplane, with $H_g$ the gas scale height. The prefactor $f_d = 0.55$ accounts for the shift of the representative size as compared to the maximum attainable size of the dust grains \citep{Bi2012}. 
   
\subsection{\label{sec:level2}Transport Model}

We assume that all the species are simultaneously released into the disk, and that this mixture has the composition of the PSN, a mixture which radially diffuses and advects. The species are initially both in solid or gaseous forms, depending on where they are in the disk.  $\Sigma_i$ represents the surface density of species $i$ we intend to study either in vaporous or solid form. We integrate the advection-diffusion equation using a forward Euler integration \citep{Bi2012} :

\begin{equation}
\frac{\Sigma_i}{\partial t} + \frac{1}{r}\frac{\partial}{\partial r} \Bigg[ r\Bigg(\Sigma_i v_i - D_i\Sigma_g\frac{\partial}{\partial r}\Bigg(\frac{\Sigma_i}{\Sigma_g}\Bigg)\Bigg)\Bigg] + \dot{Q} = 0
\end{equation}
     
\noindent where $v_i$ and $D_i$ are the radial velocities and the diffusivities of species $i$ respectively. $\dot{Q}$ corresponds to the source term of species $i$ vapor released to the gas and is given by $\Sigma_\text{ice}/\Delta t$ beyond the iceline, with~$\Delta t$~the timestep of the simulation. The last values to calculate are the velocity of the dust $v_d$, the Stokes number St, the radial velocity of the gas, and the diffusivity \citep{Bi2012}. $v_d$ is given by:

\begin{equation}
v_d = -\frac{2 \text{St}}{1+\text{St}^2}\eta v_K + \frac{1}{1+\text{St}^2}v_g,
\end{equation}

\noindent where $v_g$ is the radial inward gas velocity given by $v_g = -3\nu/2r$, with $\nu$ the turbulent diffusivity of the gas described in Sec. \ref{sec:2.1}, and $\eta$~=~$-(1/2)h^2_g d\ln P/d\ln r$ is a measure of the pressure support of the disk, where $h_g \equiv H_g/r$ is the aspect ratio of the disk. In other words, the parameter $\eta$ describes the deviation of the azimuthal velocity of the gas from the keplerian value, $v_{\phi,g}=(1-\eta) v_K$. The Stokes Number describes the aerodynamic properties of the particles and is determined as follows:

\begin{equation}
\text{St} = \frac{a\pi\rho_s}{2\Sigma_g}.
\end{equation}

Finally, the diffusivity $D$ of the vapor species is assumed to be that of the gas $D_g = \nu$ and the diffusivity of the dust is given by:
\begin{equation}
D_d = \frac{D_g}{1+\text{St}^2}.
\end{equation}

\subsection{\label{sec:level2}Source term}

The dust surface density decreases in the CPD until almost no particles are left over very short time frames (fewer than $10^4$ yrs). A disk that depletes at this rate would not be able to survive over the timescales necessary for satellite formation. Our observations on the sustainability of particles in Saturn's CPD agree with those of \cite{Ro17}. Their model for the motion of particles of different sizes evolving in the Jupiter system model found that small particles in the order of 10$^{-6}$ m moved very rapidly in Jupiter's disk, often dispersing entirely in less than 500 years. Particles over 10$^{-1}$ m ablated in much less time, often entirely sublimating in less than 20 years. 

These simulations show that no disk could be sustainable at the rate of accretion we are seeing here. An additional source of solids needs to be added so that our solid particles could continue to exist over the timescales necessary for the formation of moons. To overcome this issue, we follow the prescription of \cite{Ca2002} that injects solids directly into the disk at the centrifugal radius $R_c$. As gas and solids are delivered to the disk, the gas then sustains a quasi-steady state, while the surface density of the solids build up over time. While the gas component of the disk viscously spreads outward and onto the planet, the solids rapidly accumulate in the region where they are initially delivered, providing a mechanism for accreting large satellites in a limited region extending from the surface of the planet to the centrifugal radius $R_c$. This would also explain why we do not see large moons beyond this radius, despite the tidal stability of the region. The injection rate of solids $\dot{\Sigma}_\text{solids}$ at the centrifugal radius $R_c$ of the CPD is then \citep{Ca2002}:

\begin{equation}
\dot{\Sigma}_\text{solids} =
\begin{cases}
\frac{f\times F_p}{\pi R_c^2}  &  [r<R_c] \\
0    & [r>R_c]
\end{cases},
\end{equation}

\noindent where $f$ is the solids-to-gas ratio in the region of regular satellites, here assumed to be $\sim$7.5 $\times$ 10$^{-4}$ (volume mixing ratio). This value corresponds to the sum of the gas phase abundances of the species considered in our study (see Table 1).

\begin{figure}[b]
\includegraphics[width=\linewidth]{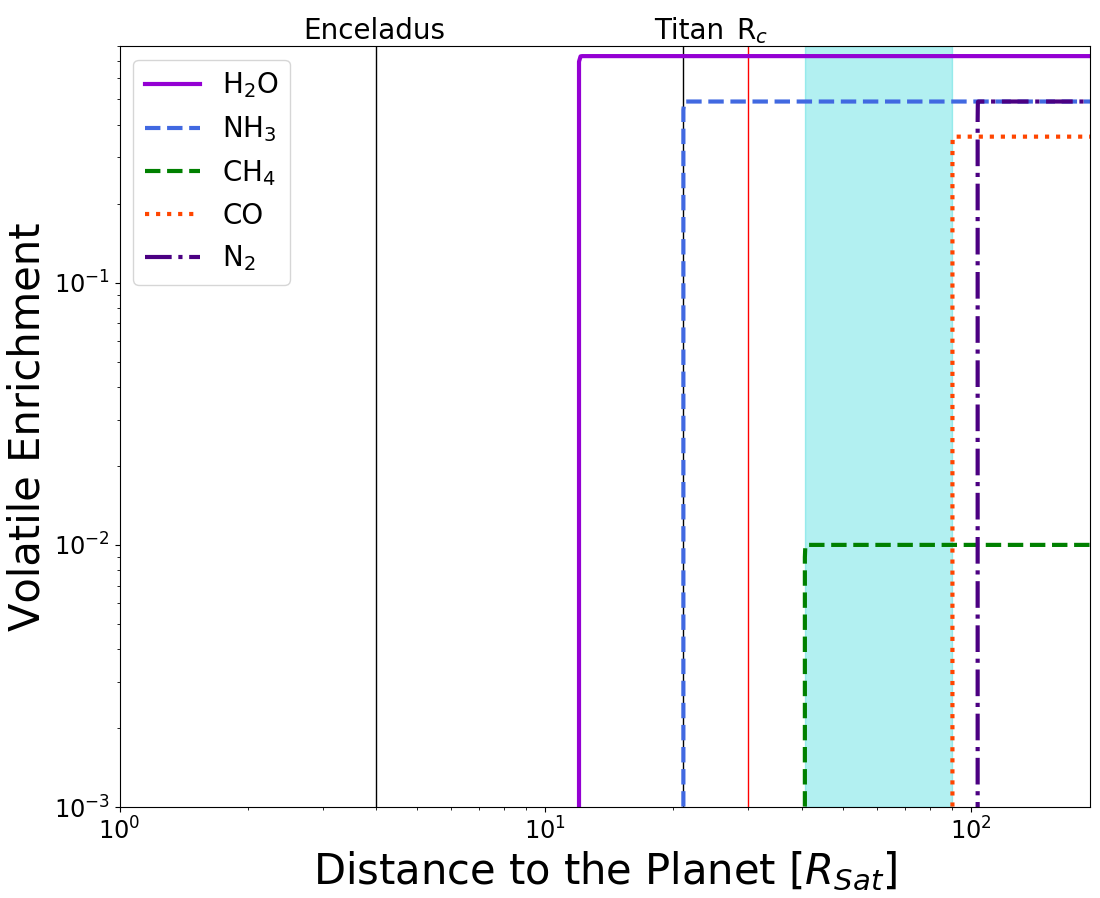}\label{fig2}
\caption{Initial enrichments in water, ammonia, methane, carbon monoxide, and nitrogen in solid forms, scaled to the elemental abundances, assumed to be protosolar \citep{Lo03}. Current orbits of Enceladus and Titan, as well as the location of the centrifugal radius $R_c$, are shown for reference. The blue rectangle represents the ideal location for the formation of the building blocks of both moons in order to match their observed compositional data.}
\end{figure}

\begin{table}[h]
\begin{center}
\footnotesize
\caption{Parameters adopted for equations (see Sec. \ref{sec:model}) describing the equilibrium vapor pressure curves of different condensates. Gas Phase abundances are provided as volume mixing ratios relative to H$_2$.}
\label{tab:dist}
\begin{tabular}{lccc}
\hline
\hline
Species X 	& A 			& B 			& X/H$_2$			\\
\hline
H$_2$O 	& -1750.286 	& 7.2326 		& $4.43 \times 10^{-4}$		\\
NH$_3$ 	& -2878.23	& 8.00205 	& $4.05 \times 10^{-5}$		\\
CH$_4$ 	& -2161.81 	& 11.1249 	& $3.16 \times 10^{-6}$		\\
CO 		& -411.24 		& 5.2426 		& $2.21 \times 10^{-4}$		\\
N$_2$ 	& -360.07 		& 4.7459 		& $4.05 \times 10^{-5}$		\\
\hline
\end{tabular}
\end{center}
\end{table}   

\section{\label{sec:res}Results}

\begin{figure*}[ht]
\includegraphics[width=\linewidth]{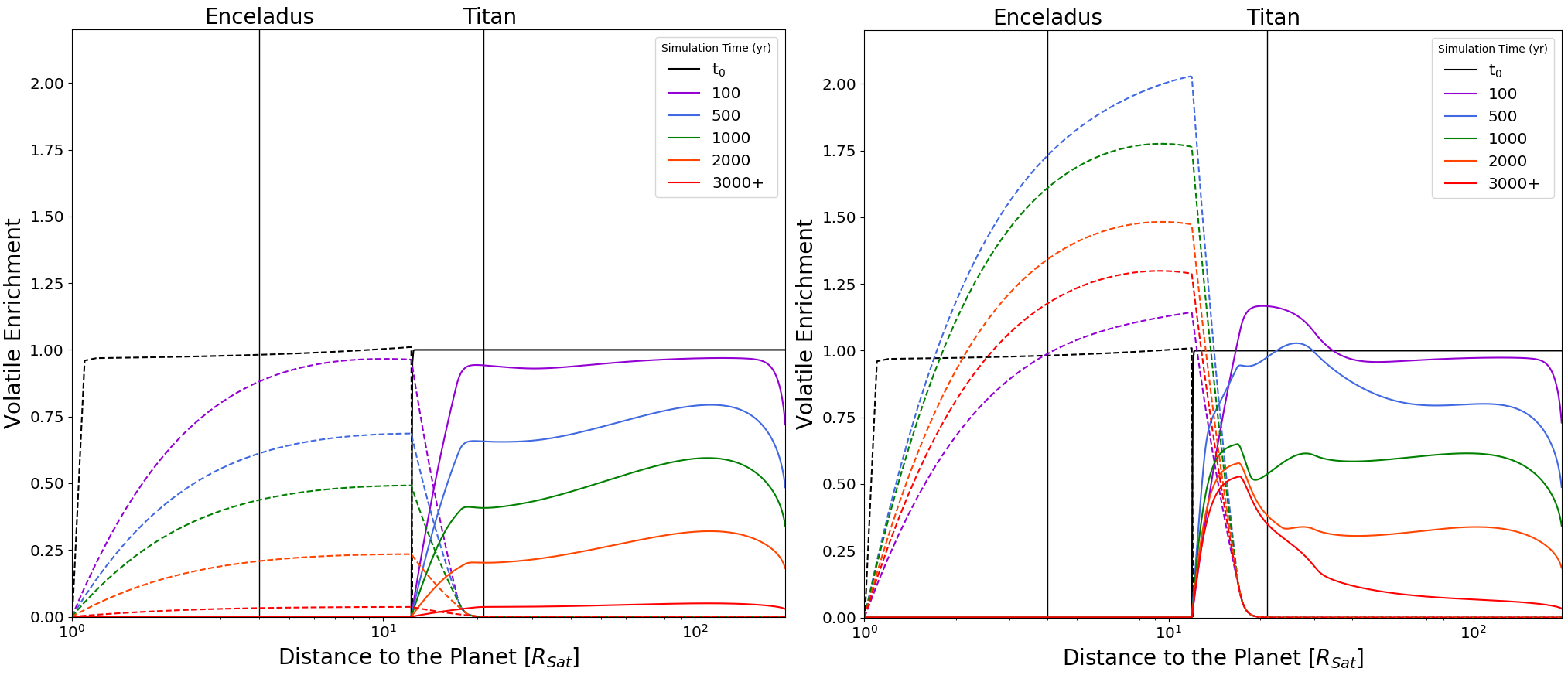}
\caption{Evolution of water vapor (dashed lines) and ice (solid lines) in the Saturnian system over 5000 yr of evolution as a function of distance to Saturn without injection of solids (left) and with the injection of solids (right). The dust and the vapor are normalized to the initial abundance of water (H$_2$O/H$_2$) = 4.43 $\times 10^{-4}$. Two things are evident: first, how quickly the particles move in the system. Second, how quickly the disk depletes: after a few thousand years, almost no dust or vapor remain if there is no source of solids.}
\label{fig3}
\end{figure*}

We place our work in the context of the satellite formation scenario proposed by \cite{Mo09a,Mo09b}, in which the building blocks of both Titan and Enceladus are assumed to be formed between the locations of the CH$_4$ clathrate iceline and those of CO and N$_2$ pure condensates in Saturn's CPD. By doing so, the satellites building blocks are presumed to be devoid in primordial CO and N$_2$ while they retain the trapped CH$_4$, in agreement with Titan and Enceladus' composition measurements.

In order to determine the positions of the icelines of H$_2$O ice, NH$_3$ hydrate, CH$_4$ clathrate, CO, and N$_2$ pure condensates, we compute the partial pressure of species $i$ at each incremental radius. We then calculate the equilibrium pressure of the species based on the temperature at that radius and compare it to the partial pressure. Equilibrium pressure equations for H$_2$O, CO, and N$_2$ pure ices derive from \cite{Mo08} and are in the form $\log P$~=~A/$T$ + B, where $P$ and $T$ are the partial pressure (bars) and the temperature (K) of the considered species, respectively. Equilibrium pressure equations for NH$_3$ hydrate and CH$_4$ clathrate derive from \cite{He04}, and are in the form $\ln P$ = A/$T$ + B. If the partial pressure of the species exceeds its equilibrium pressure at that radius, the species will be solid. Otherwise, it will sublimate. The gas phase composition of Saturn's CPD is also assumed to be protosolar \citep{Mo09a}. The abundances of the different species are derived from the assumption that all O is distributed between H$_2$O and CO, all N is in the form of N$_2$ and NH$_3$, and C only in the form of CO and CH$_4$, with a CH$_4$/CO ratio of 0.014 \citep{Mo09a}. Table 1 shows the parameters A and B derived from \cite{He04} and \cite{Mo08} and adopted for the different equilibrium curves, as well as the abundances of relevant species taken from \cite{Mo09a}. In the following, we first investigate the evolution of solids within the CPD assuming the location of $R_c$ is 30~$R_{Sat}$, a value consistent with those estimated in Jupiter's CPD \citep{Ca2002}, then we study the influence of placing this parameter to higher saturnocentric distances.

Each species is characterized by its own surface density in the CPD, either in gas or solid phase, whose value is derived from the disk's surface density and its individual abundance. At $t$ = 0, the CPD model is filled with dust whose composition is determined from our initial ratios. Any species located at closer distances to Saturn than its corresponding iceline is in gaseous phase. Beyond their respective icelines, these species are in dust form with sizes of 10$^{-6}$ m. 

\begin{figure}[t!]
\includegraphics[width=\linewidth]{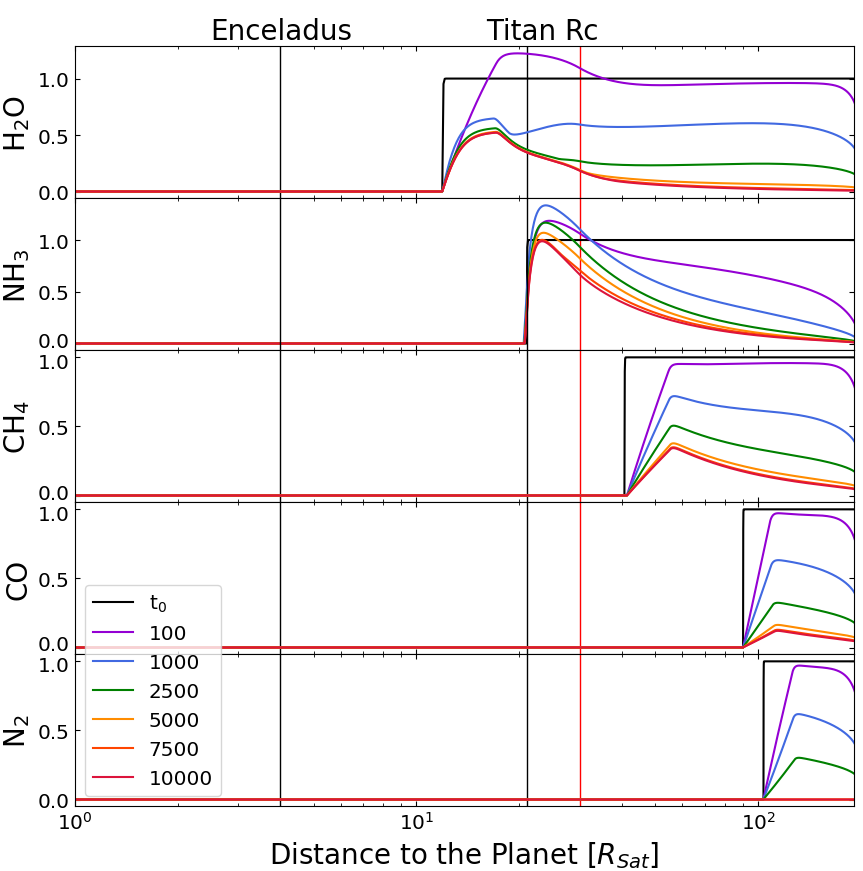}
\caption{Evolution of H$_2$O, NH$_3$, CH$_4$, CO, and N$_2$ dust normalized to their initial abundances (see Table 1) in the Saturnian system over $10^4$ years of evolution as a function of distance to Saturn with the injection of solids. Species whose condensation radii are close to $R_c$ (here NH$_3$, and H$_2$O to a lesser extent) keep significant abundances while the others (CH$_4$, CO, and N$_2$) deplete with time.}
\label{fig4}
\end{figure}

\subsection{Evolution of solids assuming $R_c$~=~30~$R_{Sat}$}

Figure 2 represents the quantity of dust in the system before any motion has occurred, giving us a visual representation of the location of icelines. At this epoch, the sublimation temperatures and iceline locations are 153.5 K and 12.0 $R_{Sat}$ for H$_2$O, 96.5K and 21.1 $R_{Sat}$ for NH$_3$ hydrate, 53.4 K and 40.7 $R_{Sat}$ for CH$_4$ clathrate, 22.4 K and 90.4 $R_{Sat}$ for CO, and 20.3 K and 103.7 $R_{Sat}$ for N$_2$ pure condensates.

As the disk cools with time, the icelines migrate inward toward Saturn. However, on the short timescales that we are interested in, the disk can be seen as stationary and the inward migration of the icelines ignored. If Titan and Enceladus mainly assembled from pebbles formed between the iceline of CH$_4$ clathrate and those of CO and N$_2$ pure condensates, as it is investigated here, the very slow cooling of our CPD model implies that the two moons never formed at their current positions. While it is possible for us to adjust the parameters of the disk so as to fit this constraint, there is no firm indication that the building blocks of the two bodies actually formed at their current location (see Sec. \ref{sec:dis}). They may have migrated in from further out in the disk, which is what we assume here. 

As evident in Fig. 3, the dust surface density decreases over very short time frames in the CPD until none is left. In the case of H$_2$O, there is no gas or dust remaining in the disk after a mere 5 $\times$ $10^3$ years. This would seem to indicate that the moons would have formed very quickly, which is impossible based on even the most optimistic estimates, which require approximately $10^4$ years \citep{Ca18}. As a result, we are forced to add a source term as established in section 2.4, where a steady injection of solids is set at $R_c$. As we inject solids into the disk, most will continue to be sublimated as they are pulled through the different icelines during their inward drift. Figure \ref{fig3} shows that, at a given epoch of the CPD evolution, the solid water abundance grows and then decreases radially because of the diffusion of vapor outward the iceline. At any given location of $R_c$, the water abundance also increases and then decreases over time until it reaches a steady state. This state is reached after 3~$\times~10^3$ years. This creates a zone where the quantity of solid H$_2$O is stable for a long period of time. 

Figure 4 depicts the evolution of H$_2$O, NH$_3$, CH$_4$, CO, and N$_2$ in solid phases as a function of the distance to Saturn in the CPD over $10^4$ years of disk evolution, with the injection of solids. Because the icelines of NH$_3$ and (to a lesser extent) of H$_2$O are closer to the location of the injection point of solids ($R_c$), these species keep significant solid abundances in the disk. CO, CH$_4$, and N$_2$, whose respective icelines are far beyond the position of $R_c$, are depleted before enough time has passed for any building blocks to form. Also, altering the solids-to-gas ratio has no effect on the lifetime of these solids as it has no influence on the location of injection. In this case, the building blocks of Titan and Enceladus cannot incorporate significant amounts of CH$_4$ from the CPD to explain their compositional properties. Here, the only possible alternative is to assume a secondary origin for CH$_4$ in these two bodies (see Sec. \ref{sec:dis}).

\begin{figure*}[t]
\includegraphics[width=\linewidth]{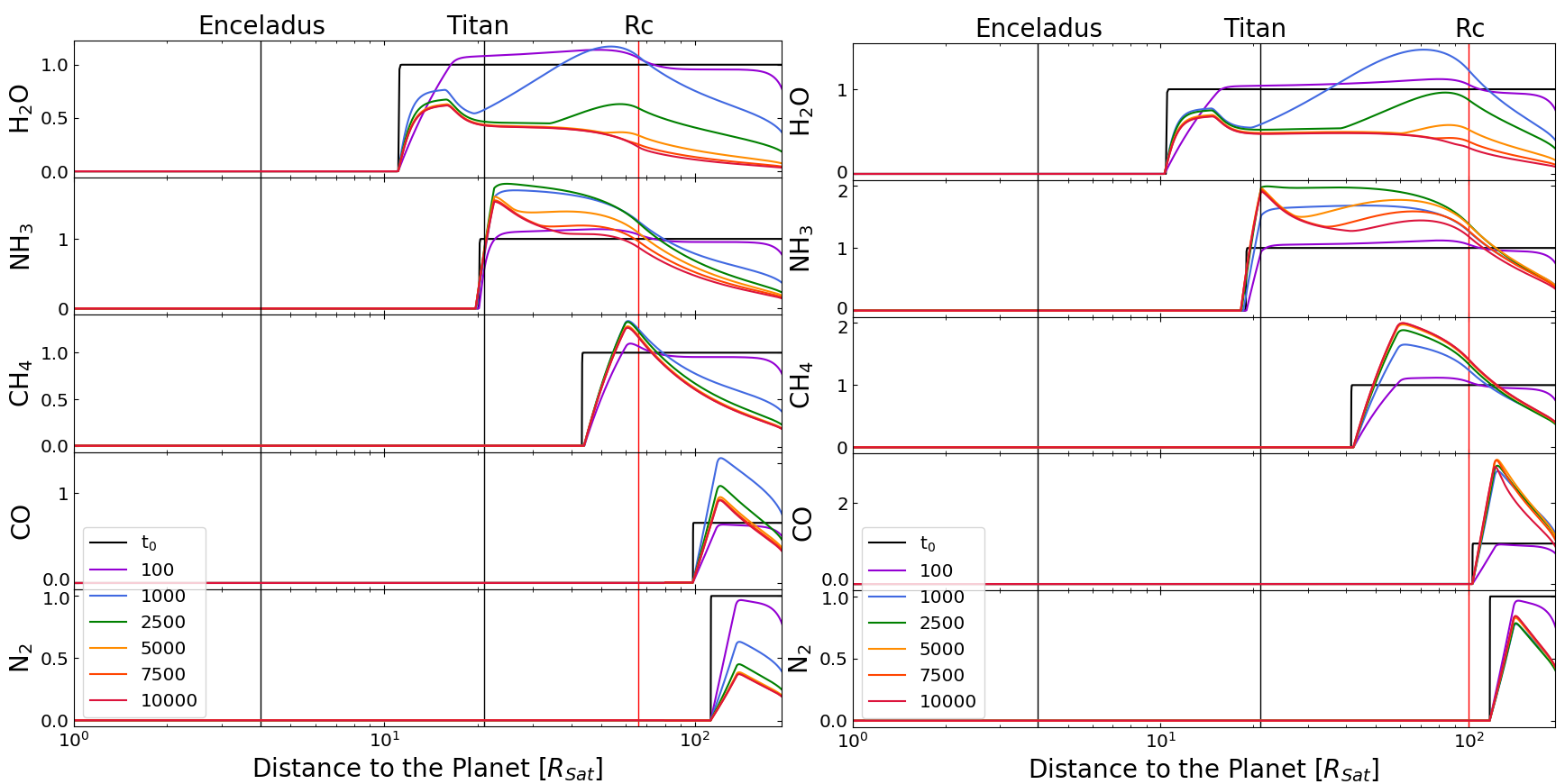}
\caption{Evolution of H$_2$O, NH$_3$, CH$_4$, CO, and N$_2$ dust normalized to their initial abundances (see Table 1) in the Saturnian system over $10^4$ years of evolution as a function of distance to Saturn with the injection of solids at $R_c$= 66$ R_{sat}$ (left) and $R_c$= 100$~R_{sat}$ (right). The most consistent position of the centrifugal radius would be in between these two values. If placed beyond, the carbon monoxide solids would be highly abundant, in contrast with the inference that Titan and Enceladus accreted from CO- and N$_2$-depleted building blocks.}
\label{fig5}
\end{figure*}

\subsection{Evolution of solids assuming $R_c > 30 R_{Sat}$}

 In this section we examine the influence of the position of $R_c$ on the solid abundances in the CPD. The distance out to which material is accreted onto the CPD remains uncertain and could be much larger than the current radial extent of the satellite systems of the giant planets \citep{Du18}. In order to form the building blocks of the Saturnian moons, we would ideally need to be at a position at which the abundance of dust significantly exceeds its initial abundance, based on the minimum conditions to develop streaming instability and planetesimals \citep{YJC17}. By expanding the centrifugal radius to larger distances, we are able to inject solids nearer each species' iceline and allow for the formation and growth of solids at higher radii.  As a result, the species are able to evolve in the system for longer periods of time, even reaching an equilibrium. 

Figure \ref{fig5} shows simulations similar to those represented in Fig. \ref{fig4}, but for $R_c$~=~66$R_{sat}$ and 100$R_{sat}$, the former value corresponding to the centrifugal radius found by \cite{Ma09}. Beneath $R_c$ = 66 $R_{sat}$, we are unable to form and maintain solids for long periods of time. Above $R_c$~=~100 $R_{sat}$, the abundances of CO and N$_2$ are too high for current observations. For the moons to form within the constraints of the estimation of the primordial composition of their building blocks, $R_c$ would have had to be within this interval, which is also bracketed by the positions of the CH$_4$ clathrate and CO icelines. Within this interval, the abundances of primordial N$_2$ and CO dust are negligible while those of H$_2$O, NH$_3$, and CH$_4$ are significant, leading to a composition similar to that expected for the building blocks of Titan and Enceladus.

Interestingly, the variation of the location of the centrifugal radius $R_c$ affects both the temperature profile of the CPD and the positions of the various icelines. Species with icelines located at distances interior to $R_c$ (such as H$_2$O and NH$_3$ hydrate) progress inwards with $R_c$ moving outward. In contrast, species whose icelines lie outside of $R_c$ (such as CO and N$_2$ pure condensates) gradually move backwards. Table 2 summarizes the locations of the various icelines in Saturn's CPD as a function of the position of $R_c$.

\begin{table}[h]
\label{tab2}
\begin{center}
\footnotesize
\caption{Positions of the icelines as a function of the value of $R_c$ (in units of $R_{sat}$).}
\label{tab:dist}
\begin{tabular}{lccc}
\hline
\hline
Species 	& 30$R_c$ 			& 66$R_c$ 			& 100$R_c$			\\
\hline
H$_2$O 		& 12.0 			& 11.2  					& 10.5				\\
NH$_3$ 		& 21.1			& 20.2 					& 18.9				\\
CH$_4$ 		& 40.7			& 43.4					& 41.6				\\
CO 			& 90.4  			& 98.5  					& 102.9				\\
N$_2$ 		& 103.7			& 112.4					& 117.1				\\
\hline
\end{tabular}
\end{center}
\end{table}  

\section{Discussion and Conclusion}
\label{sec:dis}

By using a classical prescription for Saturn's CPD, we have investigated the time evolution of the icelines of H$_2$O ice, NH$_3$ hydrate, CH$_4$ clathrate, CO, and N$_2$ pure condensates, as well as their impact on the formation conditions of the building blocks of Titan and Enceladus. To match their compositional data, the building blocks of both moons would have had to form in a region of the CPD located between the icelines of CO and N$_2$ at their outer limit, and the iceline of CH$_4$ clathrate as their inner limit. We find however that a source of dust at the location of $R_c$ does not guarantee the replenishment of the disk in the volatiles assumed to be primordial in Titan and Enceladus. 

The centrifugal radius was initially envisioned to roughly match the radial extent of the current satellite systems, with typically assumed values of $R_c$~$\sim$~20--30~$R_{Sat}$, so as to account for their compactness relative to the expected sizes of the CPDs \citep{Ru82,Ca2002,Ca2006}. In this case, we show that only the abundance of solid water remains substantial irrespective of time because its iceline is inside the centrifugal radius $R_c$. Any volatile species whose iceline lies beyond depletes too rapidly for any planetesimal formation to occur. We also performed simulations for $R_c$ values of 66 and 100~$R_{Sat}$, considering the studies made by \cite{Ma09} and \cite{Su17} regarding the structure of Saturn's CPD, and still assuming that the injection point of matter is at the location of $R_c$. By doing so, we are able to inject solids nearer each species’ iceline and allow for the formation and growth of solids with compositions consistent with those measured in Enceladus and Titan at radii between 66 and 100~$R_{Sat}$. As a result, the species are able to evolve in solid forms in the system for longer periods of time, even reaching an equilibrium, thus favoring the formation of Titan and Enceladus' building blocks in this region of the CPD.  Our results suggest that the dynamical evolution of the CPD matters little, at least to some extent. Indeed, one is able to predict the location of $R_c$ and moon building block formation solely based on the initial pressure-temperature profile of the disk.

Our results also imply that a Saturn's CPD presenting a large centrifugal radius is also consistent with the formation of Iapetus, which is currently located at $\sim$59~$R_{Sat}$, because matter remains on longer timescales in the outer CPD. Interestingly, because the clathration temperature of Xe  is higher than that of CH$_4$, this noble gas should be incorporated as well in the satellites building blocks, implying that an alternative explanation is needed to explain its deficiency in Titan's atmosphere \citep{Mo09a}. Several post-formation scenarios have already been proposed, including the removal of Titan's noble gases by their sequestration in surface clathrates \citep{Mo11}, or their trapping by the haze present in the atmosphere \citep{Ja08}. One can note that the Cassini INMS did not confirm the absence of Xe shown by the Huygens GCMS because this element was beyond the mass range of the instrument \citep{Wa09}.

The dust internal density was set to $\rho_s~=~1\text{g cm}^{-3}$ in all our calculations, corresponding to a pure icy composition in the absence of porosity. To assess the influence of the combined roles of porosity and density, we performed dust evolution simulations for $\rho_s$ varying between 0.5 and 2 $\text{g cm}^{-3}$. In all cases, the results are qualitatively similar to those obtained with $\rho_s = 1\text{g cm}^{-3}$ and do not alter our conclusions.

There remain many uncertainties regarding the formation mechanism of Saturn's moon system. Some scenarios envision that all the moons accreted within the gaseous CPD of the planet \citep{ME03a,ME03b,Ca2006}. In this case, a massive moon such as Titan could have migrated inward over large distances within the CPD through tidal interaction with the gas disk \citep{Ca2002,Ca2006,Ro20,FO20}. Titan could have thus formed at much larger orbital distances than its present location, in line with our findings regarding the composition of solids in the CPD. Titan's migration could either have been stranded close to its present orbit due to migration traps arising from strong thermal gradients in the CPD \citep{FO20} or, alternatively, it could have migrated to much closer distances from Saturn and subsequently migrated outwards due to tidal interaction with the planet \citep{LCF+20}.

In the case of Enceladus, its much smaller mass precludes a scenario involving the migration of the moon over substantial distances. However, several scenarios propose that the small and mid-sized moons of Saturn are not primordial but rather represent a second generation of satellites which would either derive from the spreading of material from Saturn's ring \citep{CCC+11,CC12,SC17}, or from the disruption of a primordial system consisting of larger moons, akin to the Galilean system \citep{SG12,AR13}. In the latter scenario, it is possible that the primordial moons were massive enough to have migrated over substantial distances in the gaseous CPD during their accretion. If, on the other hand, Enceladus accreted from material deriving from Saturn's ring, the relevance of our results depends on the origin of the rings, which remains highly debated \citep[see,][for a recent review]{Ida19}. \citet{Canup10} proposed that the rings could have originated from the tidal disruption of a massive (comparable to Titan) moon that would have formed in the gaseous CPD and migrated interior to the Roche radius of the planet. This scenario would remain consistent with the hypothesis of primordial methane accreted by the forming Titan and Enceladus, and is supported by the findings of the Cassini INMS instrument which identified CH$_4$, CO$_2$, CO, N$_2$, H$_2$O, NH$_3$, and organics in the D ring material during the Grand Finale \citep{Mi20}. 

The accretion of primordial CH$_4$ by the moons, as proposed here, implies that the subdisk was fed in low-temperature solids originating from the nebula, which were containing methane.  Our model works only in the case where the inward drift of grains and particles does not exceed the metric size. However, the bulk of the solid material could have been delivered to the CPD through the capture and ablation of planetesimals on initially heliocentric orbits \citep{Mo10, Ro18, Ro20}. Small planetesimals ($r~<10$ km) could thus contribute a significant amount of material in the outer regions of the CPD. Such planetesimals would be CO- and N$_2$-rich and their large sizes would prevent the significant loss of ultravolatiles during their migration inward the CPD, due to their low thermal conductivity \citep{Ro17}.

On the other hand, the presence of liquid water in the interiors of Titan and Enceladus \citep{Ie12,Wa17} could convert CO either into CO$_2$ or carbonate (depending on pH) for oxidizing conditions, or reduce it to organic carbon and, eventually, CH$_4$ \citep{Sh93,Gl18}. There is also evidence from the bulk composition of comet 67P/Churyumov-Gerasimeko that ice-poor outer solar system building blocks may exist \citep{Ch20}. This process may also explain the depletion of CO in Titan's atmosphere relative to atmospheric N$_2$ and CH$_4$. It may also produce post-accretion N$_2$ from either NH$_3$ and/or organic nitrogen \citep{Gl09,Ne17}. In other words, aqueous processing of primordial volatiles in satellite oceans could also generate compositions similar to those inferred in the atmosphere of Titan or in the plumes of Enceladus. A way to constrain the origin of CH$_4$ in the plumes of Enceladus would be the measurement of its D/H ratio. If the D/H ratio in CH$_4$ is close to the measurement of D/H in H$_2$O made by the INMS instrument aboard the Cassini spacecraft ($\sim$2.9 $\times$ 10$^{-4}$; \cite{Wa09}), then this methane should be the outcome of hydrothermal reactions \citep{Mo09b}. In contrast, a substantially lower value would be compatible with a primordial origin, i.e. methane originating from the PSN \citep{Mo09b}. Data on D/H in cometary methane would provide a complementary test. One would also expect a net depletion in $^{36}$Ar in the plumes of Enceladus, similarly to Titan's atmosphere, if the two moons formed following our scenario. On the other hand, the plume may deplete the interior of Enceladus in volatiles. A low $^{36}$Ar abundance could instead reflect a prolonged history of plume outgassing.

Our prescription of the CPD is quite rudimentary so the positions of the icelines are only indicative, since they are relative to our model, and different initial temperatures for the CPD alter the location of these lines. However, our model remains consistent with the Cassini data for a large range of $R_c$. 

\acknowledgements

We thank two anonymous Referees for their very useful remarks and comments. O.M. acknowledges support from CNES.

\end{document}